\documentclass[letterpaper,prl,twocolumn,showpacs,superscriptaddress]{revtex4}
\usepackage{graphicx}
\usepackage{times}

\begin{document}

\title{Experimental demonstration of continuous variable polarization
entanglement}
\author{Warwick P. Bowen}
\author{Nicolas Treps}
\author{Roman Schnabel}
\author{Ping Koy Lam}
\affiliation{Department of Physics, Faculty of Science, Australian
National University, ACT 0200, Australia}
\begin{abstract}
We report the experimental transformation of quadrature entanglement
between two optical beams into continuous variable polarization
entanglement.  We extend the inseparability criterion proposed by Duan
{\it et al.} \cite{Duan00} to polarization states and use it to
quantify the entanglement between the three Stokes operators of the
beams.  We propose an extension to this scheme utilizing two
quadrature entangled pairs for which all three Stokes operators
between a pair of beams are entangled.
\end{abstract}
\maketitle
The polarization state of light has been extensively studied in the
quantum mechanical regime of single (or few) photons.  The
demonstration of entanglement of the polarization states of pairs of
photons has been of particular interest.  This entanglement has
facilitated the study of many interesting quantum phenomena such as
Bell's inequality \cite{Aspect82}.  Comparatively, research on
continuous variable quantum polarization states has been cursory. 
Recently, however, interest in the field has increased due to the
demonstration of transfer of continuous variable quantum information
from optical polarization states to the spin state of atomic ensembles
\cite{Hald99}; and to its potential for local oscillator-free
continuous variable quantum communication networks.  A number of
theoretical papers have now been published \cite{General, Ralph00}, of
particular interest is the work of Korolkova {\it et al.}
\cite{Korolkova01} which introduces the new concept of continuous
variable polarization entanglement, and proposes methods for its
generation and characterization.  Previous to the work presented here
however, only the squeezing of polarization states had been
experimentally demonstrated~\cite{Grangier87,Hald99,Bowen02}.

In this paper we report the experimental transformation of the
commonly studied and well understood entanglement between the phase
and amplitude quadratures of two beams (quadrature entanglement)
\cite{Ou92} onto a polarization basis.  Quadrature entanglement can be
characterized using the inseparability criterion proposed by Duan {\it
et al.} \cite{Duan00}.  We generalize this criterion to an arbitrary
pair of observables and apply it to the Stokes operators that define
quantum polarization states.  We experimentally generate entanglement
of Stokes operators between a pair of beams, satisfying both the
inseparability criterion, and the product of conditional variances
which is a signature of the EPR paradox \cite{Reid88}.  Interacting
this entanglement with a pair of distant atomic ensembles could
entangle the atomic spin states.  We also analyze the polarization
state generated by combining two quadrature entangled pairs.  We show
that if the quadrature entanglement is strong enough to beat a bound
$\sqrt{3}$ times lower than that for the inseparability criterion,
then all three Stokes operators can be simultaneously entangled.

The polarization state of a light beam can be described as a Stokes
vector on a Poincar\'{e} sphere and is determined by the four Stokes
operators \cite{JauchRohrlich76}: $\hat S_{0}$ represents the beam
intensity whereas $\hat S_{1}$, $\hat S_{2}$, and $\hat S_{3}$
characterize its polarization and form a cartesian axis system.  If
the Stokes vector points in the direction of $\hat S_{1}$, $\hat
S_{2}$, or $\hat S_{3}$, the polarized part of the beam is
horizontally, linearly at 45$^\circ$, or right-circularly polarized,
respectively.  Quasi-monochromatic laser light is almost completely
polarized, in this case $\hat S_{0}$ is a redundant parameter,
determined by the other three operators.  All four Stokes operators
can be measured with simple experiments \cite{Korolkova01}.  Following
\cite{JauchRohrlich76} we expand the Stokes operators in terms of the
annihilation $\hat a$ and creation $\hat a^{\dagger}$ operators of the
constituent horizontally (subscript H) and vertically (subscript V)
polarized modes
\begin{eqnarray}
\label{stokes}
\hat S_{0}& \!= \hat a_{H}^{\dagger} \hat a^{ }_{H} + \hat a_{V}^{\dagger}
\hat a_{V} ~,\hspace{2mm}
\hat S_{2}& \!=
\hat a_{H}^{\dagger} \hat a_{V}^{ } e^{i\theta} \!+ \hat a_{V}^{\dagger}
\hat a_{H}^{ } e^{-i\theta} ,\;\;\;\;\;\;\\\nonumber
\hat S_{1}& \!= \hat a_{H}^{\dagger} \hat
a_{H}^{ } - \hat a_{V}^{\dagger} \hat a_{V}^{ } ~,\hspace{2mm}
\hat S_{3}& \!=i\hat
a_{V}^{\dagger} \hat a_{H}^{ } e^{-i\theta} \!-i\hat a_{H}^{\dagger} \hat
a_{V}^{ } e^{i\theta} ,\;\;\;\;\;\;
\end{eqnarray}
where $\theta$ is the phase difference between the H,V-polarization
modes.  Eqs.~(\ref{stokes}) are an example of the well known bosonic
representation of angular momentum type operators in terms of a pair
of quantum harmonic oscillators introduced by
Schwinger~\cite{Schwinger65}.  The commutation relations of the
annihilation and creation operators $[ \hat a_{k} \!  , \!  \hat
a_{l}^{\dagger}] \!  = \!  \delta_{kl}$ with $k,l \!  \in \!\{H,V\}$
directly result in Stokes operator commutation relations,
\begin{equation}
[\hat S_{1}, \hat S_{2}] =  2 i \hat S_{3} ~,\hspace{2mm}
[\hat S_{2}, \hat S_{3}] =  2 i \hat S_{1} ~,\hspace{2mm}
[\hat S_{3}, \hat S_{1}] =  2 i \hat S_{2} ~\,.
\label{Scomrel}
\end{equation}
These commutation relations dictate uncertainty relations between the
Stokes operators which indicate that entanglement is possible between
the Stokes operators of two beams, we term this continuous variables
polarization entanglement.  Three observables are involved, compared
to two for quadrature entanglement, and the entanglement between two
of them relies on the mean value of the third.  To provide a proper
definition of this entanglement, we have chosen to extend the
inseparability criterion proposed by Duan {\it et al.}~\cite{Duan00}.
\begin{figure}[b]
   \centerline{\includegraphics[width=7cm]{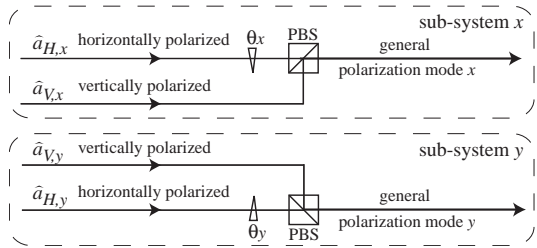}}
     \vspace{0mm} \caption{Production of arbitrary polarization modes.
     PBS: polarizing beam splitter.}
     \label{Theory}
\end{figure}
The inseparability criterion characterizes the separability of, and
therefore the entanglement between, the amplitude $\hat X \!  ^{+}$
and phase $\hat X^{-}$ quadratures of a pair of optical beams (denoted
throughout by the subscripts $x$ and $y$) with Gaussian noise
statistics.  These quadrature operators are observables and can be
obtained from the annihilation and creation operators, $\hat X \! 
^{+} \!  = \!  \hat a \!  + \!  \hat a^{\dagger}$, $\hat X \!  ^{-} \! 
= i(\!  \hat a^{\dagger} \!  - \!  \hat a)$.  In this paper we
restrict ourselves to the symmetric situation where all experimental
outcomes are independent of exchange of beams $x$ and $y$, in this
case the inseparability criterion can be written as
\begin{equation}
\label{Duan}
\Delta \!  _{x \pm y}^{2} \!  \hat X \!  ^{+} + \Delta \!  _{x \pm y}^{2} \!
\hat X \!  ^{-} < 4
\end{equation}
Throughout this paper $\Delta \!  ^{2} \!  \hat O \!  = \!  \langle
\delta \hat O ^{2} \rangle$ where $\hat O \!  = \!  \langle \hat O
\rangle \!  + \!  \delta \hat O$.  $\Delta \!  _{x \pm y}^{2} \!  \hat
O$ is the smaller of the sum and difference variances of the operator
$\hat O$ between beams $x$ and $y$, $\Delta \!  _{x \pm y}^{2} \! 
\hat O \!  = \!  \min{\langle ( \delta \hat O_{x} \!  \pm \!  \delta
\hat O_{y})^{2} \rangle}$.  Note that for physically realistic
entanglement between two observables, one observable must be
correlated, and the other anticorrelated between subsystems $x$ and
$y$.  The minimization utilized in calculating $\Delta \!  _{x \pm
y}^{2} \!  \hat O$ selects the relevant sign for each observable.  The
measure in eq.~(\ref{Duan}) relies explicitly on the uncertainty
relation between the amplitude and phase quadrature operators.  Given
the general Heisenberg uncertainty relation $ \Delta\!^{2} \hat A
\Delta\!^{2} \hat B \!  \ge \!  |\langle \delta \hat A \delta \hat B
\rangle |^{2} \!  = \!  | [\delta \hat A, \delta \hat B] |^{2}/4 \!  +
\!  |\langle \delta \hat A \delta \hat B \!  + \!  \delta \hat B
\delta \hat A \rangle|^{2}/4$ \cite{Haus00} it can be generalized to
any pair of observables $\hat A$, $\hat B$.  Unlike the commutation
relation $|[\delta \hat A, \delta \hat B] |$, the correlation function
$|\langle \delta \hat A \delta \hat B \!  + \!  \delta \hat B \delta
\hat A \rangle|$ is state dependant.  In this work we assume it to be
zero and arrive at the sufficient condition for inseparability
\begin{equation}
\label{DuanGeneral}
\Delta \!  _{x \pm y}^{2} \!  \hat A + \Delta \!  _{x \pm y}^{2} \! 
\hat B < 2 |[\delta \hat A, \delta \hat B] |
\end{equation}
To allow direct analysis of our experimental results, we define the
degree of inseparability $I(\hat A,\hat B)$, normalized such that
$I(\hat A,\hat B) \!  < \!  1$ guarantees the state is inseparable
\begin{equation}
\label{insep}
I(\hat A,\hat B)=\frac{\Delta\!  _{x \pm y}^{2}\!  \hat A+\Delta\! 
_{x \pm y}^{2}\!  \hat B} {2 |[\delta \hat A, \delta \hat B] |}
\end{equation}

An arbitrary pair of polarization modes may be constructed by
combining horizontally and vertically polarized modes on a pair of
polarizing beam splitters as shown in fig.~\ref{Theory}.  In the
symmetric situation, which this paper is restricted to, the
horizontally (vertically) polarized input beams must be
interchangeable; therefore, their expectation values and variances
must be the same ($\alpha_{H}^{ } \!  = \!  \langle \hat a_{H,x}^{ }
\rangle \!  = \!  \langle \hat a_{H,y}^{ } \rangle$, $\alpha_{V}^{ }
\!  = \!  \langle \hat a_{V,x}^{ } \rangle \!  = \!  \langle \hat
a_{V,y}^{ } \rangle$, $\Delta \!  ^{2} \!  \hat X^{\pm}_{H} \!  = \! 
\Delta\!^{2} \hat X_{H,x}^{\pm} \!  = \!  \Delta\!^{2} \hat
X_{H,y}^{\pm}$, $\Delta \!  ^{2} \!  \hat X^{\pm}_{V} \!  = \! 
\Delta\!^{2} \hat X_{V,x}^{\pm} \!  = \!  \Delta\!^{2} \hat
X_{V,y}^{\pm}$), and the relative phase between horizontally and
vertically polarized modes for subsytems $x$ and $y$ must be related
by $\theta \!  = \!  \theta_{x} \!  = \!  \pm \theta_{y} \!  + \!  m
\pi$ where $m$ is an integer.  Given these assumptions it is possible
to calculate $I(\hat S_{i}, \hat S_{j})$ from eqs.~(\ref{stokes}) and
(\ref{Scomrel}).  We choose to simplify the situation further,
providing results that may be directly related to our experiment.  We
assume that the horizontal and vertical inputs are not correlated, and
that each input beam does not exhibit internal amplitude/phase
quadrature correlations.  Finally we assume that the vertically
polarized input modes are bright ($\alpha_{V}^{2} \!  \gg \!  1$) so
that second order terms are negligible.  The denominators of
eq.~(\ref{insep}) for the three possible combinations of Stokes
operators are then found to be
\begin{eqnarray}
\label{DuanStokesS1S2S3}
|[\delta\hat S_{1}\delta, \hat S_{2}]| &=& 4 \alpha_{H} \alpha_{V}
\sin \!  {\theta} \nonumber \\
|[\delta\hat S_{1}, \delta\hat S_{3}]|   &=&  
4 \alpha_{H} \alpha_{V} \cos \!  {\theta}  \\
|[\delta\hat S_{2}, \delta\hat S_{3}]| &=& 2 | \alpha_{H}^{2} -
\alpha_{V}^{2} | \nonumber
\end{eqnarray}
In our experiment the phase $\theta$ between the horizontally and
vertically polarized input modes was controlled to be $\pi/2$, in this
situation $|[\delta\hat S_{1}, \delta\hat S_{3}]| \! = \! 0$ which 
means that using the inseparability criterion of eq.~(\ref{insep}) it is 
impossible to verify entanglement between $\hat S_{1}$ and $\hat 
S_{3}$.   On the other hand  $|[\delta\hat S_{1}, \delta\hat S_{2}]|$ 
and  $|[\delta\hat S_{2}, \delta\hat S_{3}]|$ both have finite values 
and therefore the potential for entanglement. 
%
We experimentally determined $I(\hat S_{1}, \hat S_{2})$ and $I(\hat
S_{2}, \hat S_{3})$ from measurements of $\alpha_{V}$, $\alpha_{H}$,
and $\Delta \!  _{x \pm y}^{2} \!  \hat S_{i}$.

The experimental transformation between quadrature and polarization
entanglement demonstrated here becomes clearer if $\Delta \!  _{x \pm
y}^{2} \!  \hat S_{i}$ are expressed in terms of quadrature operators. 
Assuming that $\alpha_{H}^{2} \!  \ll \!  \alpha_{V}^{2}$ we find from
eqs.~(\ref{stokes}) that $\Delta \!  _{x \pm y}^{2} \!  \hat S_{1} \! 
= \!  \alpha_{V}^{2} \Delta \!  _{x \pm y}^{2} \!  \hat X \! 
_{V}^{+}$, $\Delta \!  _{x \pm y}^{2} \!  \hat S_{2} \!  = \! 
\alpha_{V}^{2} \Delta \!  _{x \pm y}^{2} \hat X \!  _{H}^{-}$, and
$\Delta \!  _{x \pm y}^{2} \!  \hat S_{3} \!  = \!  \alpha_{V}^{2}
\Delta \!  _{x \pm y}^{2} \!  \hat X \!  _{H}^{+}$.  $I(\hat
S_{1},\hat S_{2})$ and $I(\hat S_{2},\hat S_{3})$ can then be written
\begin{eqnarray}
I(\hat S_{1},\hat S_{2}) & = &
\frac{\alpha_{V}}{\alpha_{H}}\left(\frac{\Delta \!  _{x \pm y}^{2} \! 
\hat X \!  _{V}^{+}+\Delta \!  _{x \pm y}^{2} \hat X \! 
_{H}^{-}}{8}\right) \label{insepS1S2} \\
I(\hat S_{2},\hat S_{3}) & = & \left (1+
\frac{\alpha_{H}^{2}}{\alpha_{V}^{2}} \right) \left(
\frac{\Delta \!  _{x \pm y}^{2} \!  \hat X \!  _{H}^{+}+\Delta \!  _{x
\pm y}^{2} \hat X \!  _{H}^{-}}{4} \right)
\label{insepS2S3}
\end{eqnarray}
Eq.~(\ref{insepS1S2}) shows that as $\alpha_{V}/\alpha_{H}$ increases
the level of correlation required for $I(\hat S_{1},\hat S_{2})$ to
fall below unity and therefore to demonstrate inseparability quickly
becomes experimentally unachievable.  In particular, if the horizontal
inputs are vacuum states $I(\hat S_{1},\hat S_{2})$ becomes infinite
and verification of entanglement is not possible.  In contrast,
eq.~(\ref{insepS2S3}) shows that in this situation $I(\hat S_{2},\hat
S_{3})$ becomes identical to the criterion for quadrature entanglement
(eq.~(\ref{Duan})) between the two horizontally polarized inputs. 
Therefore, quadrature entanglement between the horizontally polarized
inputs is transformed to polarization entanglement between $\hat
S_{2}$ and $\hat S_{3}$.  In the following section we experimentally
demonstrate this transformation.  The asymmetry of these results
arises because the Stokes vector of the output mode of each polarizing
beam splitter is aligned almost exactly along $\hat S_{1}$ (since
$\alpha_{H}\!\gg\!\alpha_{V}$).  This creates an asymmetry in the
commutation relations of eq.~(\ref{Scomrel}) and a corresponding bias
in the uncertainty relations that define the inseparability criteria.

In our experiment two equal power 1064~nm amplitude squeezed beams
were produced in a pair of spatially separated optical parametric
amplifiers (OPAs).  The OPAs were optical resonators constructed from
hemilithic MgO:LiNbO$_{3}$ crystals and output couplers and are
described in detail in ref.~\cite{Bowen02}.  We combined the squeezed
beams with $\pi/2$ phase shift on a 50/50 beam splitter with
interference efficiency of 97.8\%.  The output beams exhibited the
conventional quadrature entanglement \cite{Ou92}.  We modematched each
entangled beam into a homodyne detector that provided amplitude or
phase quadrature measurements and characterized the entanglement with
the inseparability measure given in eq.~(\ref{Duan}).  We obtained the
result $I(\hat X^{+},\hat X^{-}) \!  = \!  (\Delta \!  _{x \pm y}^{2}
\!  \hat X \!  ^{+} \!  + \!  \Delta \!  _{x \pm y}^{2} \!  \hat X \! 
^{-})/4 \!  = \!  0.44$, which is well below unity.  We also
determined the product of conditional variances between the beams
($\min_{g}[{\langle ( \delta \hat X \!  ^{+}_{x} \!  + \!  g \delta
\hat X \!  ^{+}_{y})^{2} \rangle \langle ( \delta \hat X \!  ^{-}_{x}
\!  - \!  g \delta \hat X \!  ^{-}_{y})^{2} \rangle}] <1$), which was
propose by Reid and Drummond \cite{Reid88} as a signature of the EPR
paradox.  We observed a value of $0.58$ which is also well below
unity.

\begin{figure}[b]
   \centerline{\includegraphics[width=8.7cm]{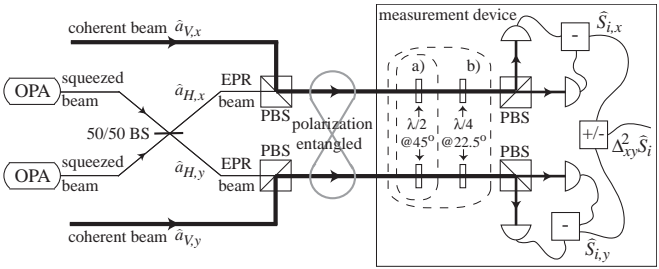}}
     \vspace{0mm} \caption{Experimental production and characterization
     of continuous variable polarization entanglement.  The optics
     within a) are included to measure $\hat S_{2}$, and those within
     b) to measure $\hat S_{3}$. (P)BS: (polarizing) beam splitter.}
     \label{experiment}
\end{figure}
We transformed the entanglement onto a polarization basis by combining
each entangled beam (horizontally polarized) with a much more intense
vertically polarized coherent beam ($\alpha_{V}^{2} \!  = \!  30
\alpha_{H}^{2}$) with measured mode-matching efficiency for both of
91\% (see fig.~\ref{experiment}).  The relative phase between the
horizontal and vertical input modes $\theta$ was controlled to be
$\pi/2$.  The two resultant beams were polarization entangled.  We
verified this entanglement by measuring correlations of the Stokes
operators between the beams.

Each beam was split on a polarizing beam splitter and the two outputs
were detected on a pair of high quantum efficiency photodiodes. 
Dependent on the inclusion of wave plates before the polarizing beam
splitter, the difference photocurrent between the two photodiodes
yielded instantaneous values for $\hat S_{1}$, $\hat S_{2}$, or $\hat
S_{3}$ (see fig.~\ref{experiment}).  The variance of the unity gain
electronic sum or subtraction of the Stokes operator measurements
between the polarization entangled beams was obtained in a spectrum
analyzer that had a resolution bandwidth of 300~kHz and video
bandwidth of 300~Hz.  This resulted in values for $\Delta \!  _{x \pm
y}^{2} \!  \hat S_{i}$.  All of the presented results were taken over
the sideband frequency range from 2 to 10~MHz and are the average of
ten consecutive traces.  Every trace was more than 4.5~dB above the
measurement dark noise which was taken into account.  We determined
$\alpha_{V}^{2}$ directly by blocking the horizontal modes and
measuring the power spectrum of the subtraction between the two
homodynes, this also gave $\alpha_{H}^{2}$ since the ratio
$\alpha_{V}^{2}/\alpha_{H}^{2}$ was measured to equal thirty.

\begin{figure}[t]
   \centerline{\includegraphics[width=8.7cm]{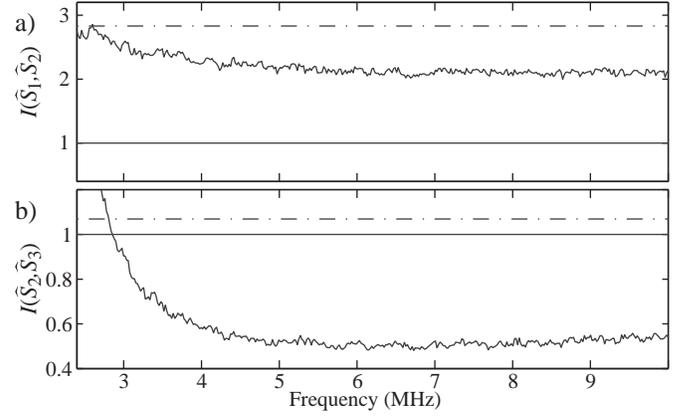}}
\vspace{0mm} \caption{Experimental measurement of a) $I(\hat S_{1},
\hat S_{2})$ and b) $I(\hat S_{2}, \hat S_{3})$, values below unity
indicate entanglement.  The dashed line is the corresponding
measurement inferred between two coherent beams.}
     \label{DuanS1S2S3}
\end{figure}
Fig.~\ref{Duan} shows our experimental measurements of $I(\hat S_{1},
\hat S_{2})$ and $I(\hat S_{2}, \hat S_{3})$.  The dashed lines
indicate the results a pair of coherent beams would produce.  Both
traces are below this line throughout almost the entire measurement
range, this is an indication that the light is in a non-classical
state.  At low frequencies both traces were degraded by noise
introduced by the relaxation oscillation of our laser.  $I(\hat S_{2},
\hat S_{3})$ shows polarization entanglement, however as expected
$I(\hat S_{1}, \hat S_{2})$ is far above unity.  The best entanglement
was observed at 6.8 MHz with $I(\hat S_{2}, \hat S_{3}) \!  = \! 
0.49$ which is well below unity.

By electronically adding or subtracting the Stokes operator
measurements with a gain $g$ chosen to minimize the resulting variance
we observed a signature of the EPR paradox for polarization states. 
In this case the product of the conditional variances of $\hat S_{2}$
and $\hat S_{3}$ from one beam after utilizing information gained
through measurement of the other must be less than the Heisenberg
uncertainty product ($\min_{g}[{\langle ( \delta \hat S \!  _{2,x} \! 
\pm \!  g \delta \hat S_{2,y})^{2} \rangle \langle ( \delta \hat S \! 
_{3,x} \!  \pm \!  g \delta \hat S_{3,y})^{2} \rangle}] \!  < \! 
|[\delta\hat S_{2}, \delta\hat S_{3}]|^{2}/4$).  We observed a
conditional variance product of $0.77 |[\delta\hat
S_{2}, \delta\hat S_{3}]|^{2}/4$.  Therefore both the degree of
inseparability and the conditional variance product demonstrate that
our experiment generates strong polarization entanglement.

\begin{figure}[b]
   \centerline{\includegraphics[width=8.7cm]{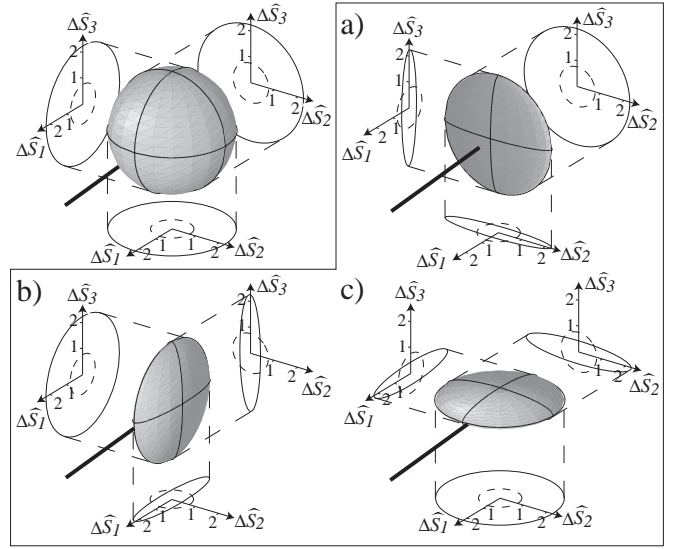}}
     \vspace{0mm} \caption{Calculated polarization entanglement
     produced from four pure quadrature squeezed beams with squeezed
     quadrature variances of 0.1; axes normalized to 1 for a coherent
     state.  The top left figure represents the knowledge of beam $y$
     before any measurement of beam $x$.  a), b), and c) represent the
     conditional knowledge of beam $y$ given measurements of $\hat
     S_{1}$, $\hat S_{2}$, and $\hat S_{3}$ respectively on beam $x$. 
     If the conditional knowledge is better than the dashed circles
     the state is entangled.}
     \label{SymmetricEPR}
\end{figure}

Polarization entanglement has more degrees of freedom than quadrature
entanglement because three observables, rather than two, are involved. 
In the following section we consider the continuous variable situation
most analogous to single photon polarization entanglement where the
correlation is independent of the basis of measurement, and
demonstrate theoretically that all three Stokes operators can be
simultaneously entangled.  We extend the work of
ref.~\cite{Korolkova01}, and arrange the entanglement such that
eqs.~(\ref{DuanStokesS1S2S3}) are equal and the mean value of the
three Stokes operators are the same ($| \!  \langle \hat S_{i}\rangle
\!  | \!  = \!  \alpha^{2}$).
This leads to $\alpha_{V}^{2} \!  = \!  \frac{\sqrt{3} -
1}{2}\alpha^{2}$, $\alpha_{H}^{2} \!  = \!  \frac{\sqrt{3} +
1}{2}\alpha^{2}$, $\theta_{x} \!  = \!  \pi/4 \!  + \!  n_{x}\pi/2$,
and $\theta_{y} \!  = \!  \pi/4 \!  + \!  n_{y}\pi/2$ where $n_{x}$
and $n_{y}$ are integers.  We assume that the two horizontally
polarized inputs, and the two vertically polarized inputs, are
quadrature entangled with the same degree of correlation such that
$\Delta \!  _{x \pm y}^{2} \!  \hat X \!  _{H}^{\pm} \!  = \!  \Delta
\!  _{x \pm y}^{2} \!  \hat X \!  _{V}^{\pm} \!  = \!\Delta \!  _{x
\pm y}^{2} \!  \hat X$.  In this configuration,
eqs.~(\ref{DuanStokesS1S2S3}) become $|\langle\delta\hat
S_{i}\delta\hat S_{j}\rangle| \!  = \!  \alpha^{2}$, for all $i \! 
\neq \!  j$.  To simultaneously minimize all three degrees of Stokes
operator inseparability ($I(\hat S_{i}, \hat S_{j})$) it is necessary
that $\theta_{x} \!  = \!  -\theta_{y} \!  + \!  n\pi$.  After making
this assumption we find that $\Delta \!  _{x \pm y}^{2} \!  \hat S_{i}
\!  = \!  \sqrt{3}\alpha^{2}\Delta \!  _{x \pm y}^{2} \!  \hat X$ for
all $i$.  Hence, in this situation $I(\hat S_{i}, \hat S_{j})$ are all
identical, and the entanglement is equivalent between any two Stokes
operators.  The condition for entanglement can then be expressed as a
simple criterion on the quadrature entanglement between the input
beams
\begin{equation}
   I(\hat S_{i}, \hat S_{j})<1 \Longleftrightarrow
   I(\hat X^{+},\hat X^{-}) < 1/{\sqrt{3}}
\end{equation}
where $I(\hat X^{+},\hat X^{-})\!=\!I(\hat X^{+}_{H},\hat
X^{-}_{H})\!=\!I(\hat X^{+}_{V},\hat X^{-}_{V})$.  The factor of
$\frac{1}{\sqrt{3}}$ arises from the projection of the quadrature
properties onto a polarization basis in which the Stokes vector is
pointing at equal angle ($\cos^{-1} (\frac{1}{\sqrt{3}})$) from all
three Stokes operator axes.  In principle it is possible to have all
the three Stokes operators perfectly entangled.  In other word,
ideally the measurement of any Stokes operator of one of the beams
could allow the exact prediction of that Stokes operator from the
other beam (see fig.~\ref{SymmetricEPR}).  The experimental production
of such a field is a straightforward extension of the experiment
reported here, given the availability of four independent squeezed
beams.  Maximal single photon polarization entanglement enables tests
of Bells inequality\cite{Aspect82}.  It has recently been shown that
continuous variable polarization entanglement of the form discussed
above can also exhibit Bell-like correlations\cite{Ralph00}.  This
entanglement resource would also enable the demonstration of maximal
continuous variable polarization teleportation.

To conclude, we have presented the first generation of continuous
variable polarization entanglement.  The scheme presented transforms
the well understood quadrature entanglement to a polarization basis. 
The two Stokes operators orthogonal to the Stokes vectors of the
polarization entangled beams easily fulfill a generalized version of
the inseparability criterion proposed by Duan {\it et al.}.  We have
also demonstrated that in the limiting case of our experimental
configuration where $\alpha_{V}^{2}\!\gg\!1$ and $\alpha_{H}^{2} \!  =
\!  0$ it is not possible to verify entanglement between any other
pair of Stokes operators.  Finally we have shown that using four
squeezed beams it is possible for all three Stokes operators to be
perfectly entangled, although with a bound $\sqrt{3}$ times lower
(stronger) than that for quadrature entanglement.

This work was supported by the Australian Research Council and is part
of the EU QIPC Project, No.  IST-1999-13071 (QUICOV).  R.~S.
acknowledges the Alexander von Humboldt foundation for support.  We are
grateful of H.~A.~Bachor and T.~C.~Ralph for invaluable discussion.

\end{document}